\documentclass[aps,prl,twocolumn,showpacs,byrevtex]{revtex4}
\let\oldhat\hat
\renewcommand{\hat}[1]{\oldhat{\mathbf{#1}}}

\usepackage{times,mathptm}
\usepackage[dvips]{graphicx,color}

\begin{document}
\title{Competing Magnetic Orderings and Tunable Topological States in Two-Dimensional Hexagonal Organometallic Lattices}
\author{Hyun-Jung Kim$^{1,2}$, Chaokai Li$^{3,4}$, Ji Feng$^{3,4{\dagger}}$, Jun-Hyung Cho$^{1*}$, and Zhenyu Zhang$^{2}$}
\affiliation{$^1$ Department of Physics and Research Institute for Natural Sciences, Hanyang University,
17 Haengdang-Dong, Seongdong-Ku, Seoul 133-791, Korea \\
$^2$International Center for Quantum Design of Functional Materials (ICQD), Hefei National Laboratory for Physical Sciences at the Microscale (HFNL), and Synergetic Innovation Center of Quantum Information and Quantum Physics, University of Science and Technology of China, Hefei, Anhui 230026, China \\
$^3$International Center for Quantum Materials, Peking University, Beijing 100871, China \\
$^4$Collaborative Innovation Center of Quantum Matter, Beijing 100871,  China
}
\date{\today}

\begin{abstract}
The exploration of topological states is of significant fundamental and practical importance in contemporary condensed matter physics, for which the extension to two-dimensional (2D) organometallic systems is particularly attractive. Using first-principles calculations, we show that a 2D hexagonal triphenyl-lead lattice composed of only main group elements is susceptible to a magnetic instability, characterized by a considerably more stable antiferromagnetic (AFM) insulating state rather than the topologically nontrivial quantum spin Hall state proposed recently. Even though this AFM phase is topologically trivial, it possesses an intricate emergent degree of freedom, defined by the product of spin and valley indices, leading to Berry curvature-induced spin and valley currents under electron or hole doping. Furthermore, such a trivial band insulator can be tuned into a topologically nontrivial matter by the application of an out-of-plane electric field, which destroys the AFM order, favoring instead ferrimagnetic spin ordering and a quantum anomalous Hall state with a non-zero topological invariant. These findings further enrich our understanding of 2D hexagonal organometallic lattices for potential applications in spintronics and valleytronics.
\end{abstract}

\maketitle


Since the discovery of the quantum Hall effect~\cite{klitzing}, the concept of topological order has become a subject of major interest in contemporary condensed matter physics~\cite{kane1}. Kane and Mele revealed~\cite{kane2} that graphene, a two-dimensional (2D) honeycomb lattice of carbon atoms~\cite{novoselov}, opens a gap by spin-orbit coupling (SOC) and realizes a topologically nontrivial state called the quantum spin Hall (QSH) state, giving rise to a quantized response of a transverse spin current to an applied in-plane electric field. This discovery triggers a huge amount of activities in exploring 2D or 3D topological insulators (TIs) that possess robust helical conducting edge or surface states on the boundary of bulk insulators~\cite{murakami,bernevig,konig,roth,zhang,xia}. This peculiar helical state is topologically protected from elastic backscattering by time-reversal symmetry~\cite{zhang,xia}, and hence offers fascinating playgrounds for applications in spintronics and quantum computation devices~\cite{hasan}.

Analogous to the quantum control of the spin~\cite{wolf,zutic}, other binary quantum degrees of freedom such as two inequivalent valleys ($K$ and $K'$ points) and two inequivalent lattice sites ($A$ and $B$ sublattices) in 2D hexagonal lattices, as shown in Fig. 1, have been exploited to bring the emergence of the so-called valleytronics~\cite{rycerz,shkolnikov,xiao,xu2} and pseudospintronics~\cite{xu2,pesin,novoselov2}. As the inversion symmetry is broken in 2D hexagonal structures, Bloch wavefunctions exhibit opposite Berry curvatures at $K$ and $K'$ to host a quantum valley Hall (QVH) effect~\cite{xiao,yao,li} characterized by a valley Chern number $C_v$ = $C_K$ $-$ $C_{K'}$. Recently, various schemes have been proposed to generate valley currents in graphene~\cite{rycerz,gunlycke,jiang} and 2D transition metal dichalcogenides~\cite{wang,zeng} by using their unique edge modes~\cite{rycerz}, defect lines~\cite{gunlycke}, and strain~\cite{jiang}. Since a strong SOC gives rise to the interplay between spin and valley and lattice pseudospins~\cite{xu2}, it is interesting and challenging to examine how it can rearrange the band structure and further change the topological property using external perturbations, e.g., electric field. Indeed, there have been several recent proposals on the topological phase transitions in 2D hexagonal structures, induced by an out-of-plane external electric field~\cite{ezawa,kim,qian}. However, the realization of these exciting possibilities in 2D materials is still in its infancy.

\begin{figure}[ht]
\includegraphics[width=0.48\textwidth]{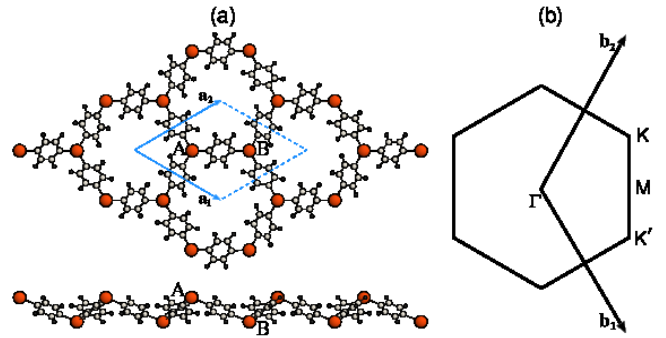}
\caption{(Color online) (a) Top and side views of the structure of the 2D hexagonal triphenyl-lead lattice. The dotted line indicates the unit cell. The large, medium, and small circles represent Pb, C, and H atoms, respectively. Two inequivalent Pb lattice sites are denoted as $A$ and $B$ sublattices. The Brillouin zone is displayed in (b), together with the high symmetric $k$-points ${\Gamma}$, $M$, $K$, and $K'$.}
\end{figure}

Here we focus on a prototypical example of 2D hexagonal organometallic systems, triphenyl-lead (TL) lattice~\cite{wang2}. As shown in Fig. 1, Pb atoms form a honeycomb lattice with a threefold rotational symmetry by sharing phenyl bridges with three neighbors, and the $A$ and $B$ sublattices of Pb atoms exhibit slight puckering. Considering that the Pb atom has the 6$s^2$6$p^2$ valence-electron configuration, one expects that the TL lattice would realize a QSH state, similar to silicene that is a puckered honeycomb lattice of Si~\cite{liu}. Indeed, in their pioneering density-functional theory (DFT) studies of topological phases in organic systems~\cite{wang2,TM}, Wang, Liu, and Liu demonstrated that the TL lattice with a nonmagnetic configuration would be a QSH insulator.

In this Letter, we perform comprehensive DFT calculations for the TL lattice to investigate its true ground state by including a quantum anomalous Hall (QAH) state as well as a band insulator with an antiferromagnetic (AFM) spin ordering. We find that the latter AFM insulating state is more stable than the QSH and QAH states. Despite its topologically trivial feature, this AFM ground state possesses a novel electronic degree of freedom defined by the product of spin and valley indices, leading to Berry curvature-induced spin and valley currents under carrier doping or polarized optical pumping. We also reveal that the application of vertical gating induces a topological phase transition from the AFM insulating state by closing one of the valley gap. As the gap reopens, the system enters into a ferrimagnetic QAH insulator with a Chern number of one. Because the AFM and QAH states with fascinating electronic properties are connected by a single experimental knob and furthermore 2D organometallic systems are chemically flexible in terms of ligands and metal centers, the generic consequences of the present findings have much to offer for the design and development of spintronics and valleytronics applications.

We begin to study the nonmagnetic (NM) state of the TL lattice using the PBE calculation without the inclusion of SOC~\cite{method}. The optimized structure is displayed in Fig. 1. The calculated distance $d_{\rm Pb-Pb}$ between two neighboring Pb atoms and their height difference ${\Delta}h$ are 7.452 and 2.932 {\AA}, respectively. Here, each phenyl ring rotates along the Pb-Pb axis by ${\sim}$27$^{\circ}$ relative to the horizontal plane of an isolated TL molecule. The PBE band structure shows Dirac cones at the $K$ and $K'$ points, with a Dirac point at the Fermi level $E_F$ (see Fig. 1S(a) of the Supplemental Material~\cite{supple}). On the other hand, when SOC is included in the PBE calculation, the band gap $E_g$ at $K$ and $K'$ is opened as high as ${\sim}$11 meV [see Fig. 2(a)], in good agreement with that (8.6 meV) of a previous DFT calculation~\cite{wang2}. By computing the parity eigenvalues of Bloch wavefunctions at the time-reversal-invariant momenta, the NM state obtained using the PBE+SOC calculation is found to be a QSH state.

\begin{figure*}[ht]
\centering{ \includegraphics[width=16.cm]{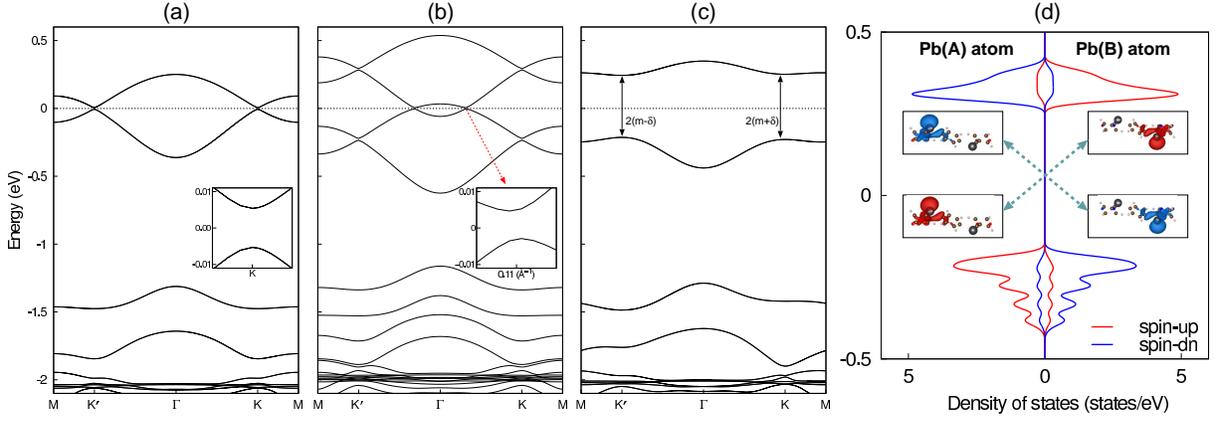} }
\caption{(Color online) Band structures of the (a) NM, (b) FM, and (c) AFM states, obtained using the PBE+SOC calculation. The energy zero represents $E_F$. The insets in (a) and (b) magnify the energy dispersion near $E_F$. For the AFM state, the PBE spin-polarized local DOS projected onto the two Pb atoms at the $A$ and $B$ sites are given in (d). The charge characters of the spin-up (spin-down) states for the highest occupied and the lowest unoccupied bands are taken at the $K$ point with an isosurface of 0.002 ($-$0.002) electrons/{\AA}$^3$}
\end{figure*}

It is noticed that Pb atoms in the puckered TL lattice favor the $sp^3$ hybridization, leaving a single dangling bond on each Pb atom. Since the Dirac bands originating from such dangling-bond electrons have a small bandwidth of ${\sim}$0.5 eV [see Fig. 2(a)], other electronic instabilities such as a charge/spin density wave (CDW/SDW) may be favored over the QSH state. For the CDW instability, we are unable to find its stabilization within the employed unit cell [see Fig. 1(a)], but the ferromagnetic (FM) and AFM states can be stabilized. The calculated PBE and PBE+SOC total energies of the FM and AFM states relative to the NM (or QSH) state are given in Table I. We find that, for the PBE (PBE+SOC) calculation, the FM state is less stable than the NM state by 6.8 (7.6) meV, while the AFM state is significantly more stable than the NM state by 37.4 (36.5) meV.

\begin{table}[ht]
\caption{Calculated total energies (in meV per unit cell) of the FM and AFM states relative to the NM state, obtained using the PBE and PBE+SOC calculations. The calculated spin magnetic moments (in ${\mu}_B$) of Pb and C atoms are also given. Here, three C atoms bonding to Pb are taken.}
\begin{ruledtabular}
\begin{tabular}{lcrr}
   		&   &  PBE  &  PBE+SOC 	 	\\  \hline
FM  &  ${\Delta}E_{\rm FM-NM}$   & 		6.8			& 7.6 						\\
    &  $M_{\rm Pb}$   &   0.35   &  0.33   \\
    &  $M_{\rm C}$    &   0.20   &  0.21   \\
AFM  &  ${\Delta}E_{\rm AFM-NM}$  	& 	$-$37.4  &  $-$36.5   \\
    &  $M_{\rm Pb}$   &   ${\pm}$0.32   &  ${\pm}$0.31   \\
    &  $M_{\rm C}$   &   ${\pm}$0.16   &  ${\pm}$0.22   \\
\end{tabular}
\end{ruledtabular}
\end{table}

Figure 2(b) shows the PBE+SOC band structure for the FM state. We find that the SOC opens a gap of $E_g$ = 7.8 meV at the two points near the ${\Gamma}$ point, while the PBE band structure shows a gap closing of the Dirac cones at $E_F$ (see Fig. 5S(b) of the Supplemental Material~\cite{supple}). From the PBE+SOC calculation, we calculate the Berry curvatures ${\Omega}$ whose integral over the Brillouin zone (see Fig. 6S of the Supplemental Material~\cite{supple}) gives the Chern number $C$ = $\frac{1}{2{\pi}} {\int}_{\rm BZ} {\Omega}~d^{2}k$ = 1, with equal contributions from the valley regions. The nonzero Chern number is the characteristic of a QAH state, similar to the recently studied triphenyl-manganese lattice~\cite{TM}.

It is remarkable that AFM order is stabilized in a organometallic compound composed of only main group elements. Contrasting with the QSH and QAH states in the TL lattice, the AFM state exhibits a large band-gap opening at valleys $K$ and $K'$ [see Fig. 2(c)]: i.e., the PBE calculation gives an identical value of $E_g$ = 469 meV, while PBE+SOC gives different values of 481 and 457 meV, respectively. This large gap opening of the AFM state gives a large thermodynamics stability relative to the NM state (see Table I). To understand the underlying mechanism for the AFM spin ordering, we plot in Fig. 2(d) the PBE spin-polarized local density of states (DOS) projected onto the Pb atoms at the A and B sublattices, together with their spin characters. It is seen that the occupied (unoccupied) spin-up and spin-down dangling-bond states are localized at the A(B) and B(A) sites, respectively. Since electronic states with the same spin direction can hybridize with each other, the hybridization takes place between the occupied and unoccupied spin-up (spin-down) states localized at the A(B) and B(A) sites, respectively, yielding not only a gain of the exchange kinetic energy but also an insulating gap~\cite{sato}. This kind of exchange interaction mediated by two electronic states which are energetically separated well below and above the Fermi level is characterized as a superexchange mechanism~\cite{super1,super2}. It is likely that such a superexchange interaction can be facilitated due to a considerable hybridization of the Pb 6$p_z$ orbitals with the ${\pi}$ orbitals of neighboring phenyl rings. As shown in Fig. 7S of the Supplemental Material~\cite{supple} and Table I, this hybridization is well represented by a large spin delocalization with the alternating spin densities of ${\sim}$${\pm}$0.32 and ${\sim}$${\pm}$0.16 ${\mu}_B$ for Pb atoms and C atoms in the phenyl ring, respectively.

In Fig. 2(c), the SOC-induced renormalization of the valley gaps in the AFM state can be described by the parameter ${\delta}$: that is, $E_{g}^{\rm SOC}$ = 2($m+{\tau}{\delta}$), where $m$ = $\frac{1}{2}$$E_{g}$ = 234.5 meV is the mass term arising from the AFM spin ordering and ${\tau}$ = 1 ($-$1) represents the valley index of $K$ ($K'$). The present value of ${\delta}$ = 6 meV in the AFM state indicates a moderate spin-valley coupling (SVC), where the product $s{\cdot}{\tau}$ of spin and valley indices becomes a new degree of freedom of electrons. Here, the spin-up and spin-down bands remain degenerate because the Hamiltonian is invariant under simultaneous time reversal $T$ and spatial inversion $P$, although neither $T$ nor $P$ alone commutes with the Hamiltonian. Figure 3(a) (Fig. 8S of the Supplemental material~\cite{supple}) shows the Berry curvatures of the lowest (highest) unoccupied (occupied) spin-up and spin-down bands, computed using PBE+SOC. It is seen that the spin-up and spin-down bands have the opposite values of ${\Omega}$ at valley $K$ and these values are also opposite to the corresponding values at $K'$, manifesting the SVC. Consequently, the Chern number becomes zero, and the insulating AFM state is characterized to be topologically trivial.

\begin{figure}[t]
\centering{ \includegraphics[width=7.7cm]{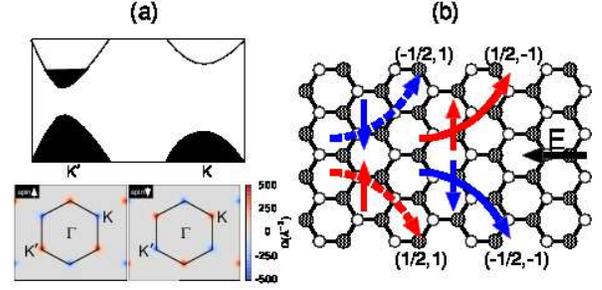} }
\caption{(Color online) (a) Schematic band structure of the AFM state with electron doping at valley $K'$ and Berry curvatures of the lowest unoccupied spin-up and spin-down bands. The spin-up and spin-down currents under an applied in-plane electric field ${\bf E}_{\rm in}$ are shown in (b), depending on the amount of doping level that shifts ${\Delta}E_F$ ${\le}$ $m$ + 2${\delta}$ and ${\Delta}E_F$ $>$ $m$ + 2${\delta}$. Solid and dashed lines stand for the currents from $K'$ and $K$, respectively. The spin and valley indices (s,${\tau}$) are indicated in parentheses.}
\end{figure}

Due to  the SVC in the AFM state, the valleys' degeneracy can be lifted without any external field~\cite{li,qi15}. For instance, the valleys can be doped asymmetrically with electrons (or holes), as schematically illustrated in Fig. 3(a). In the case of electron doping, the conduction band is occupied to shift $E_F$ to a higher energy. If this shift ${\Delta}E_F$ is smaller than $m$ + 2${\delta}$, an applied in-plane electric field produce a transversal
valley current with the electrons of ($s$,${\tau}$) = (${\pm}\frac{1}{2}$,$-$1), giving rise to a net transversal spin current without charge current, as schematically shown in Fig. 3(b). This is characterized as the QVH effect. It is notable that such spin currents are induced by the spin-dependent Berry curvatures ${\Omega}_{K'}$ [see Fig. 3(a)] that determine the anomalous velocities of Bloch electrons, V$_a$ ${\sim}$ ${\bf E}_{\rm in}$ ${\times}$ ${\Omega}_{K'}$. For ${\Delta}E_F$ $>$ $m$ + 2${\delta}$, the pair of currents with ($s$,${\tau}$) = (${\pm}\frac{1}{2}$,1) starts to contribute to a net transversal spin current in the reversal direction [see Fig. 3(b)]. When we switch from electron-doping to hole-doping, spin holes can generate valley and spin currents in the opposite directions. This generation of spin currents in the AFM state with time-reversal symmetry breaking represent the anomalous spin Hall effect.

The mean field of the present AFM state can be characterized as a spin-dependent staggered lattice potential in the honeycomb lattice. It is evident that an additional spin-independent staggered lattice potential will induce a closure of the gap in one of the valleys, eventually giving rise to a nontrivial electronic topology. To explore this topological phase transition, we apply an out-of-plane electric field ${\bf E}_{\rm out}$ to produce a staggered electric potential in the puckered honeycomb TL lattice. Here, ${\bf E}_{\rm out}$ is simulated by superimposing an additional sawtooth potential along the $z$ direction with discontinuity at the mid-plane of the vacuum region of the supercell. This staggered electric potential is expected to lift the spin degeneracy of the AFM state because the occupied spin-up and spin-down DB states are localized at the A and B sites with different heights [see Fig. 2(d)], respectively. Consequently, as the magnitude of ${\bf E}_{\rm out}$ increases, the gap of the spin-split bands would decrease, close, and reopen at the $K$ and $K'$ points, leading to a topological phase transition. Indeed, our PBE+SOC calculations show the evolution of the band structure with increasing ${\bf E}_{\rm out}$, giving rise to a gap closing of the spin-down band at valley $K'$ for the critical field of ${\bf E}_{\rm out}$ ${\approx}$ 0.62 V/{\AA}. Figure 4(a) shows the band structure and Berry curvature obtained at ${\bf E}_{\rm out}$ = 0.6 V/{\AA}. It is seen that the band gap at $K$ ($K'$) is much reduced as 119 (97) meV, compared to the value of 481 (457) meV in the absence of ${\bf E}_{\rm out}$ [see Fig. 2(c)]. The calculated spin magnetic moments for the Pb atoms at the A and B sites become $M_{\rm Pb(A)}$ = 0.28 and $M_{\rm Pb(B)}$ = $-$0.25 ${\mu}_B$, respectively, indicating a ferrimagnetic spin ordering. We note that the Chern number for this ferrimagnetic state is still zero. By contrast, for ${\bf E}_{\rm out}$ = 0.7 V/{\AA}, the Chern number is changed into 1 as a consequence of the band inversion at valley $K'$ where the sign of Berry curvature changes [see Fig. 4(b)]. Here, the spin magnetic moments for the Pb atoms are much reduced as $M_{\rm Pb(A)}$ = 0.18 and $M_{\rm Pb(B)}$ = $-$0.15 ${\mu}_B$. These results obviously indicate a phase transition to a topologically nontrivial QAH state with ferrimagnetic order. It is noticeable that, when ${\bf E}_{\rm out}$ is above ${\sim}$0.71 V/{\AA}, the gap of the spin-down band at valley $K$ closes and reopens, which makes the system back to a trivial insulator. Since 4${\delta}$ [i.e., the gap difference between $K$ and $K'$ valleys in Fig. 2(c)] amounts to 24 meV, the topological nontrivial QAH state induced by a certain out-of-plane external electric field may be realized at temperatures below ${\sim}$300 K.

\begin{figure}[ht]
\centering{ \includegraphics[width=7.7cm]{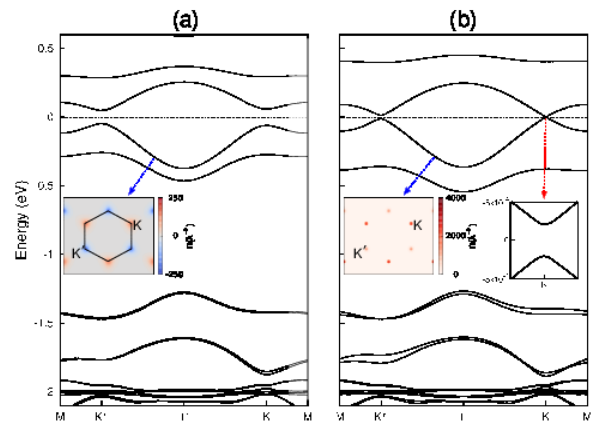} }
\caption{(Color online) Calculated band structures of (a) the topologically trivial ferrimagnetic state at ${\bf E}_{\rm out}$ = 0.6 V/{\AA} and (b) the topologically nontrivial QAH state at ${\bf E}_{\rm out}$ = 0.7 V/{\AA}. The Berry curvature of each structure is also given. }
\end{figure}

In summary, we have performed comprehensive DFT calculations for the hexagonal TL lattice to investigate its true ground state by considering the QSH and QAH states as well as a band insulator with AFM order. We find that the topologically trivial AFM band insulator is more stable than the QSH and QAH states. This AFM state gives an emergent electronic degree of freedom characterized by the product of spin and valley indices, which leads to Berry curvature-induced spin and valley currents under electron or hole doping. Furthermore, we find that the application of an out-of-plane electric field can induce a quantum phase transition from the AFM state to a topologically nontrivial QAH state with ferrimagnetic order. Our findings are fairly generic for 2D organometallic lattices, which will provide novel platforms for the applications of spintronics and valleytronics using 2D AFM hexagonal lattices. In particular, the metal centers and ligand choice in 2D organometallic lattices are two variables in their syntheses, which will surely enrich the exploration of topological and correlated electronic phases.

\vspace{0.4cm}

This work was supported in part by National Research Foundation of Korea (NRF) grant funded by the Korean Government (2015R1A2A2A01003248), the National Natural Science Foundation of China (Grants No. 11174009 and No. 61434002), and the National Key Basic Research Program of China (Grants No. 2013CB921900 and No. 2014CB921103). The calculations were performed by KISTI supercomputing center through the strategic support program (KSC-2014-C3-049) for the supercomputing application research.

\noindent $^{*}$Corresponding author: chojh@hanyang.ac.kr \\
\noindent $^{\dagger}$Corresponding author: jfeng11@pku.edu.cn


\newpage
\titlepage
\centering{\Large Supplemental Material for ``Competing Magnetic Orderings and Tunable Topological States in Two-Dimensional Hexagonal Organometallic Lattices"} \\
\vspace{1cm}
\centering{\large Hyun-Jung Kim$^{1,2}$, Chaokai Li$^{3,4}$, Ji Feng$^{3,4{\dagger}}$, Jun-Hyung Cho$^{1*}$, and Zhenyu Zhang$^{2}$}\\
\vspace{.5cm}
\centering{$^1$ Department of Physics and Research Institute for Natural Sciences, Hanyang University,
17 Haengdang-Dong, Seongdong-Ku, Seoul 133-791, Korea \\
$^2$International Center for Quantum Design of Functional Materials (ICQD), Hefei National Laboratory for Physical Sciences at the Microscale (HFNL), and Synergetic Innovation Center of Quantum Information and Quantum Physics, University of Science and Technology of China, Hefei, Anhui 230026, China \\
$^3$International Center for Quantum Materials, Peking University, Beijing 100871, China \\
$^4$Collaborative Innovation Center of Quantum Matter, Beijing 100871,  China
}

\makeatletter
\renewcommand{\fnum@figure}{\figurename ~\thefigure{S}}
\renewcommand{\fnum@table}{\tablename ~\thetable{S}}
\makeatother

\vspace{2.4cm}
{\bf \large 1. Band structures of the NM, FM, and AFM states}
\vspace{0.2cm}
\begin{figure}[h]
\centering{ \includegraphics[width=14.0cm]{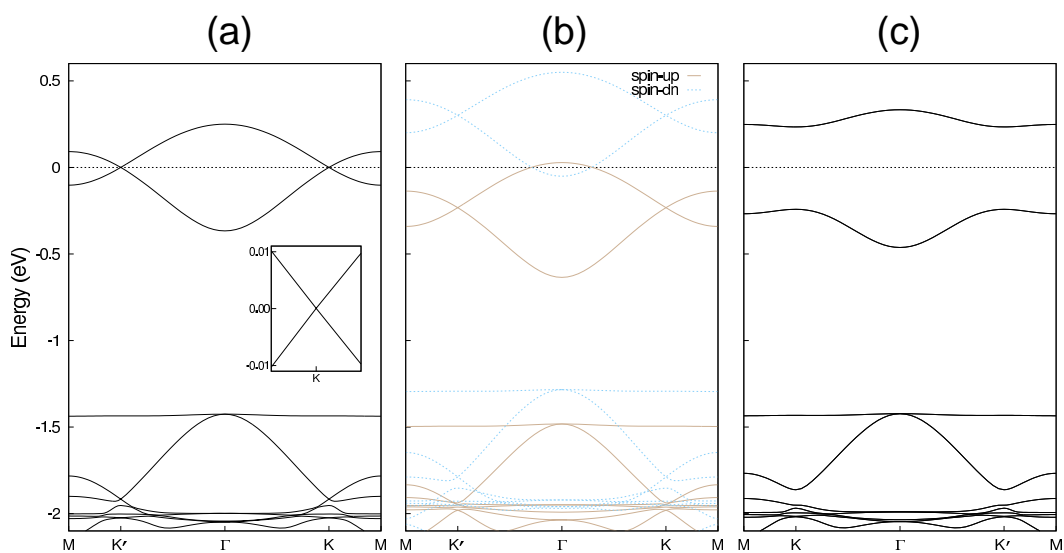} }
\caption{Calculated band structures of the (a) NM, (b) FM, and (c) AFM states, obtained using the PBE calculation. The inset in (a) magnifies the energy dispersion of the Dirac cone at the K point. In (b), the solid and dashed lines represent the spin-up and spin-down bands, respectively. The energy zero represents the Fermi level.
}
\end{figure}
\newpage
{\bf \large 2. Berry curvature of the FM state}
\vspace{0.2cm}


\begin{figure}[h]
\centering{ \includegraphics[width=14.0cm]{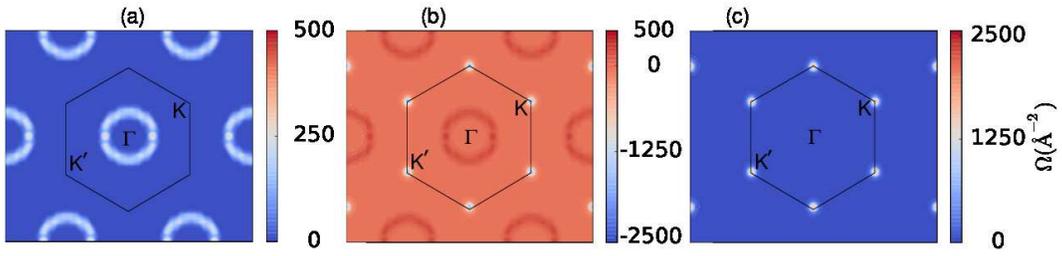} }
\caption{(a) Calculated Berry curvature for the FM state, obtained using the PBE+SOC calculation. The Berry curvatures of the highest occupied state and the second highest occupied state are also given in (b) and (c), respectively. We find that the Chern numbers calculated from (a), (b), and (c) are 1, 0, and 1, respectively.}
\end{figure}

\vspace{2cm}

\newpage
{\bf \large 3. Spin density of the AFM state}
\vspace{0.2cm}
\begin{figure}[h]
\centering{ \includegraphics[width=14.0cm]{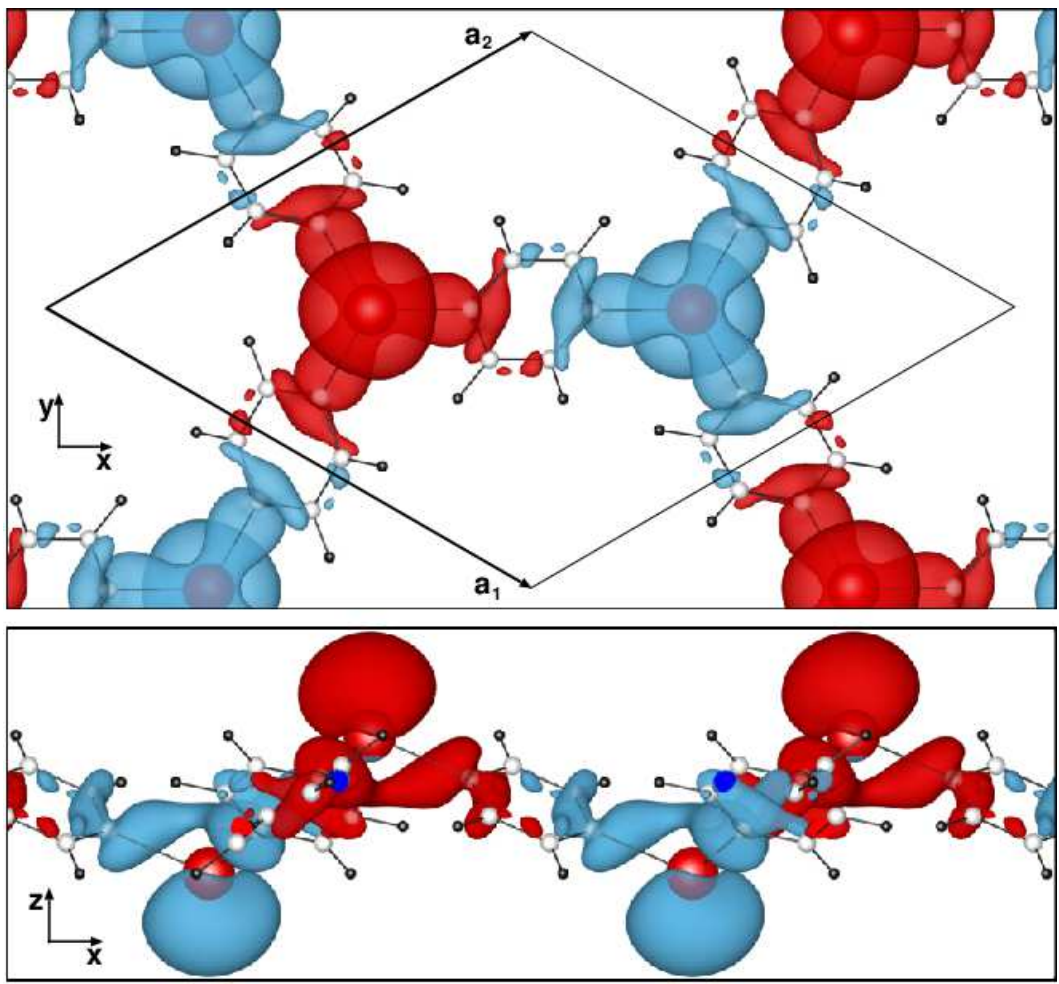} }
\caption{Tops and side views of the PBE spin density of the AFM state. The spin-up (spin-down) density is displayed in red
(blue) color with an isosurface of 0.002 ($-$0.002) $e$/\AA$^{3}$. }
\end{figure}

\newpage
{\bf \large 4. Berry curvature of the AFM state}
\vspace{0.2cm}
\begin{figure}[h]
\centering{ \includegraphics[width=14.0cm]{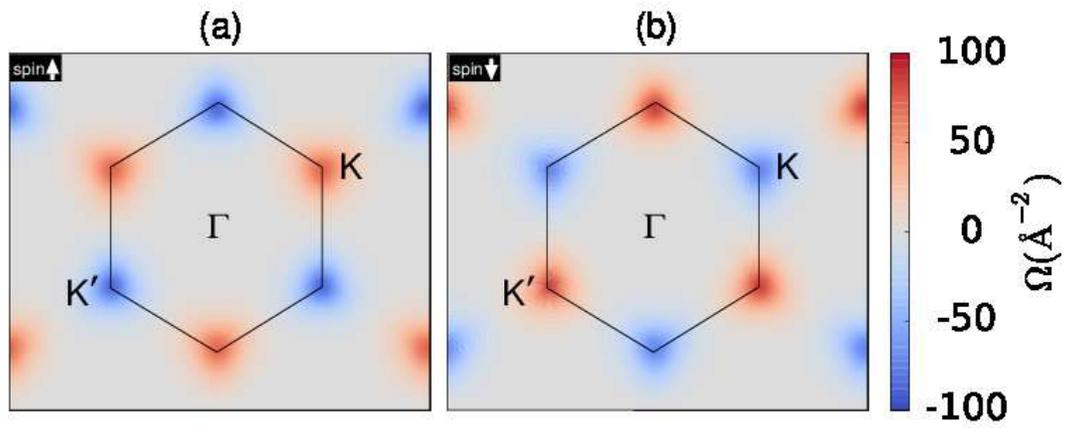} }
\caption{Calculated Berry curvatures of the highest occupied (a) spin-up and (b) spin-down bands in the AFM state, obtained using the PBE+SOC calculation.}
\end{figure}

\end{document}